\documentclass[iop,natbib]{emulateapj}

\usepackage[]{natbib}
\newcommand{\vtan}{$V_{tan}$}

\newcommand{\vrad}{$v_R$}
\newcommand{\vpec}{$v_{pec}$}
\newcommand{\vdisp}{$v_{disp}$}

\newcommand{\mura}{$\mu_{\alpha *}$}
\newcommand{\mudec}{$\mu_{\delta}$}
\newcommand{\pmra}{$\mu_{\alpha *}$}
\newcommand{\pmdec}{$\mu_{\delta}$}
\newcommand{\pmups}{$\mu_{\upsilon}$}
\newcommand{\pmtau}{$\mu_{\tau}$}
\newcommand{\kms}{km\,s$^{-1}$}
\newcommand{\masyr}{mas\,yr$^{-1}$}

\newcommand{\name}{WISE J104915.57-531906}

\slugcomment{}
\shorttitle{New Nearby Binary Brown Dwarf}
\shortauthors{Mamajek}
\begin{document}

\title{On the Nearby Binary Brown Dwarf WISE
  J104915.57-531906.1 (Luhman 16)}

\author{Eric E. Mamajek}
\affil{Department of Physics and Astronomy, 
University of Rochester, Rochester, NY 14627, USA} 
\email{emamajek@pas.rochester.edu}

\begin{abstract} 
I report some observations and calculations related to the new nearby
brown dwarf at d = 2 pc discovered by Luhman (2013, ApJ Letters, in
press; arXiv:1303.2401). I report archival astrometry and photometry
of the new object from IRAS (epoch 1983.5; IRAS Z10473-5303), AKARI
(epoch 2007.0; AKARI J1049166-531907), and the Guide Star Catalog
(epoch 1995.304; GSC2.2 S11132026703, GSC2.3 S4BM006703). A
SuperCOSMOS scan of a plate taken with the ESO Schmidt Telescope
(epoch 1984.169) shows the source as elongated (PA = 138$^{\circ}$).
Membership of the binary to any of the known nearby young groups
within 100 pc appears unlikely based on the available astrometry and
photometry. Based on the proper motion and parallax, a Monte Carlo
simulation of thin disk/thick disk/halo stars is suggestive that the
binary is, unsurprisingly, most likely a thin disk star ($\sim$96\%),
with a $\sim$4\%\, chance that it is a thick disk (and negligible
chance that it is a halo star). I propose that this important new
nearby substellar binary be called by either its provisional
Washington Double Star catalog identifier (``Luhman 16''), or perhaps
``Luhman-WISE 1'', either of which is easier to remember than the WISE
identifier.
\end{abstract}


\section{Introduction}

\citet{Luhman13} recently discovered a new system (WISE
J104915.57-531906.1) which appears to be the nearest "star"\footnote{I
  use the term ``star'' loosely, as it is clear that the components of
  the binary are clearly substellar.} to the Sun discovered in nearly
a century (Proxima Centauri was discovered in 1915, and the distance
to Barnard's star was constrained by the mid-1910s as well). This
remarkable discovery was made possible using astrometry and photometry
from the NASA WISE mission \citep{Wright10}.\\

As this system will become an important benchmark object for the study
of substellar L dwarfs, and no doubt attract follow up investigations,
I decided to present my notes on this object in a timely manner. In
\S\,2, I discuss new detections of the object in the AKARI, IRAS, and
GSC catalogs (epochs between 1983 and 2007), and a SuperCOSMOS
detection of the object on an ESO Schmidt Telescope plate (epoch
1984).  In \S\,3 I discuss the kinematics of the star, and in \S4 I
discuss the nomenclature of this interesting new system.\\

\section{Appearance in Other Catalogs}

\citet{Luhman13} found optical/IR counterparts to WISE
J104915.57-531906.1 in various astrometric/photometric catalogs, and
listed the following aliases: 2MASS J10491891-5319100 and DENIS
J104919.0-531910.  The object is fairly bright in 2MASS ($J$ = 10.73,
$H$ = 9.56, $K_s$ = 8.84), and conspicuous in the WISE All-Sky Data
Release catalog ([W1] $<$ 8, [W4] $<$ 6!).\\

Based on the 2MASS position (epoch 1999.375) and the proper motion in
Luhman (2013; \mura\, = -2759\,$\pm$\,6 \masyr, \mudec\, =
354\,$\pm$\,6 \masyr), I estimate the ICRS/J2000 position for epoch
J2000.0 to be approximately 10:49:18.723 -53:19:09.86 ($\alpha$,
$\delta$ = 162.3280125, -53.3194046). I have not yet included the
parallactic motion, so this position for epoch 2000.0 is currently
accurate to only $\sim$1". In Table \ref{tab:pos}, I list predicted
positions of the binary at various useful past and future epochs
(accuracy $\sim$1''). The coordinates are of sufficient accuracy to
search for pre-WISE astrometry in various astronomical catalogs.\\

\begin{deluxetable*}{lccccc}[htb!]
\tabletypesize{\scriptsize}
\setlength{\tabcolsep}{0.03in}
\tablewidth{0pt}
\tablecaption{Predicted Positions for WISE
J104915.57-531906.1\label{tab:pos}}
\tablehead{
{(1)}   &{(2)}      & {(3)}      &{(4)}\\
{Epoch} &{$\alpha_{ICRS}$} & {$\delta_{ICRS}$} &{Notes}        
}
\startdata
2014   & 10:49:14.41 & -53:19:05 & \\
2013   & 10:49:14.72 & -53:19:05 & \\
2010   & 10:49:15.64 & -53:19:06 & WISE epoch\\
2007.0 & 10:49:16.57 & -53:19:07 & AKARI epoch\\
2005   & 10:49:17.18 & -53:19:08 & \\
2000   & 10:49:18.72 & -53:19:10 & \\
1995.3 & 10:49:20.17 & -53:19:12 & GSC epoch\\
1995   & 10:49:20.26 & -53:19:12 & \\
1990   & 10:49:21.80 & -53:19:13 & \\
1984.169 & 10:49:23.60 & -53:19:15 & ESO Schmidt epoch\\
1983.5 & 10:49:23.80 & -53:19:16 & IRAS epoch\\
1980   & 10:49:24.88 & -53:19:17 & \\
1970   & 10:49:27.96 & -53:19:20 & \\
1960   & 10:49:31.04 & -53:19:24 & \\
1950   & 10:49:34.12 & -53:19:28 & \\
1900   & 10:49:49.51 & -53:19:45 &
\enddata
\end{deluxetable*}

\subsection{AKARI}

I tested whether the object may be the counterpart to AKARI/IRC point
source J1049166-531907, which lies 9.37" away from the WISE
coordinates (epoch $\sim$2010).  The mean epoch for the AKARI point
source position is not listed in the AKARI/IRC All-Sky Survey Point
Source Catalog Version
1.0\footnote{http://cdsarc.u-strasbg.fr/viz-bin/Cat?II/297}
\citep{Ishihara10}, however the AKARI cyrogenic mission ran between UT
8 May 2006 (2006.351) and 26 August 2007 (2007.652), hence I assume a
mean epoch for the AKARI position of 2007.0. For epoch 2007.0,
Luhman's binary brown dwarf would have been near ICRS position
10:49:16.57 -53:19:07. This position is within 1" of the position of
AKARI J1049166-531907 (epoch 2007.0), and this AKARI/IRC source is the
only one within 4.7 arcminutes. The AKARI point source has a
positional uncertainty ellipse of 0".45 $\times$ 0".21 (PA = 89.66
deg). Hence, the positional agreement between the AKARI point source
position and the predicted 2007.0 position of the new binary is
excellent. The AKARI/IRC All-Sky Survey detected this source in 9
scans, and quotes a 9\,$\mu$m flux in the AKARI/9SW band of 1.558e-1
$\pm$ 9.33e-03 Jy. No 18\,$\mu$m flux is listed, and the object is not
listed as a source in the AKARI/FIS (Far-infrared 50-180 $\mu$m)
survey. Adopting the zero-magnitude flux density of 56.26 Jy from
\citet{Ishihara10}, this flux density translates to magnitude [9] =
6.39\,$\pm$\,0.07 mag. This is similar to the magnitude quoted in the
WISE Channel 3 (11.6\,$\mu$m band) -- [W3] = 6.200\,$\pm$\,0.015 --
further securing the cross-identification.\\

\subsection{IRAS}

It is possible that Luhman's binary combined with another WISE source
(WISE J104924.57-531910.5) may together contribute infrared flux to
the IRAS Faint Source Reject Catalog source IRAS Z10473-5303. The IRAS
astrometry is for epoch J1983.5 and the IR source is at 10:49:24.6
-53:19:16, which is only 7.4" away from where Luhman's object is
predicted to be for that epoch (IRAS position uncertainties are an
ellipse of 16".6 x 5".2 at PA = 144$^{\circ}$). This was a low S/N
12\,$\mu$m detection (F$_{12}$ = 0.1657 Jy, $\pm$17.8\%). For a
Vega-like SED, this flux would translate to a 12\,$\mu$m magnitude of
5.53 ($\sim$20\% error). Luhman's binary has [W3] = 6.20 mag (with a
decidedly un-Vega-like SED), and WISE J104924.57-531910.5 has W3 =
7.44. These two objects would have been the brightest objects at
12\,$\mu$m within 1', and both likely contributed to the IR flux of
IRAS Z10473-5303.\\

\subsection{Guide Star Catalog}

A plausible optical counterpart is detected in versions 2.2 and 2.3 of
the Guide Star Catalog at epoch 1995.304. Based on the
\citet{Luhman13} proper motion, I predict that the binary should
appear at ICRS position 10:49:20.17 -53:19:12 at epoch 1995.304. The
binary appears to correspond to the optical counterpart GSC2.2
S11132026703 (ICRS 10:49:20.156 -53:19:11.02, epoch 1995.304, $R$ =
18.16\,$\pm$\,0.44, no $B_J$ or $V$ magnitude given) and GSC2.3
S4BM006703 (ICRS 10:49:20.156 -53:19:11.02, epoch 1995.304, $F$ =
18.16\,$\pm$\,0.44, $j$\,=\,22.25\,$\pm$\,0.55). The ``$F$''
photographic magnitude reported in GSC 2.3 was for IIIaF emulsion with
OG590 R-band filter.  The astrometry in the GSC 2.2 has errors of
0.419" in RA and 0.399" in Dec. The GSC2.2 and GSC 2.3 counterparts
lie within 0".5 of the predicted position, however there is another
unrelated 2MASS source 2.92" away which may be blending the
photometry/astrometry.\\

\subsection{SuperCOSMOS}

As \citet{Luhman13} mentioned, the star is visible on ``DSS IR'' and
``DSS red'' plates taken in 1978 and 1992, respectively. I checked the
SuperCOSMOS plate archive\footnote{http://www-wfau.roe.ac.uk/sss/}
\citep{Hambly01} operated by the Institute for Astronomy (University
of Edinburgh) to find their measured astrometry and magnitudes.\\

{\it UKST red (1992):} SuperCOSMOS lists the equinox J2000 position as
($\alpha$, $\delta$ = 162$^{\circ}$.3383111, -53$^{\circ}$.3202878 =
10:49:21.1947 -53:19:13.036) on the UKST red plate OR
14804\footnote{The housekeeping file for the scan can be found at:
  http://www-wfau.roe.ac.uk/sss/cgi-bin/hk.cgi?survey=UKR\&field=169.}
with IIIaF emulsion with the OG590 red filter. The epoch is listed as
1992.191, and a modified Julian Date of 48689.595616944 (which differs
by a day from that quoted in Luhman 2013). SuperCOSMOS estimated a red
magnitude of 18.578. This position differs from that measured directly
by \citet{Luhman13} by only 0''.16, well within the his quoted 0''.30
positional uncertainty.\\

{\it UKST IR (1978):} SuperCOSMOS lists the equinox J2000 position as
($\alpha$, $\delta$ = 162$^{\circ}$.3562217, -53$^{\circ}$.3216400 =
10:49:25.4932 -53:19:17.904) on the UKST infrared plate I
4176\footnote{The housekeeping file for the scan can be found at:
  http://www-wfau.roe.ac.uk/sss/cgi-bin/hk.cgi?survey=UKI\&field=169.}
with IVN emulsion and RG715 filter. The epoch is listed as 1978.303,
and a modified Julian Date of 43616.502665653 (which again differs
from that listed by Luhman 2013 at the one-day level). SuperCOSMOS
measured an IVN/RG715 magnitude of 15.289. This position differs from
the measured directly by \citet{Luhman13} by 0''.55 (where he
estimated a position uncertainty of 0''.40). As \citet{Luhman13}
commented, on the 1978 plate the object appears to be slightly blended
with a fainter star to the east.\\

{\it ESO red (1984):} This is a new detection, and perhaps the most
interesting one. SuperCOSMOS provides imagery for this region from the
ESO Schmidt Telescope, for which a red (IIIaF emulsion, RG630 filter)
image\footnote{The housekeeping file for this scan can be found at:
  http://www-wfau.roe.ac.uk/sss/cgi-bin/hk.cgi?survey=ESOR\&field=169.}
(ESO red plate R 5562) was taken epoch 1984.169 (MJD
45760.21756326). The quoted red magnitude is 18.501. Using the
\citet{Luhman13} proper motion solution, I estimate that at this
epoch, the object should appear near ICRS position 10:49:23.6
-53:19:15. An {\it elongated} point source appears near this position
at ICRS position ($\alpha$, $\delta$ = 162$^{\circ}$.3487647,
-53$^{\circ}$.3211742 = 10:49:23.7035 -53:19:16.227). The elongated
object on the plate appears in a rather empty region of sky where no
other point sources appear in any of the images shown in Fig. 1 of
\citet{Luhman13}, hence the identification of the object on the ESO
red plate with the new binary seems very secure. The SuperCOSMOS fit
to the elongated object has major axis 35.77 $\mu$m, minor axis 26.60
$\mu$m with position angle PA = 138$^{\circ}$. The position angle
appears remarkably similar to that of the resolved pair in the GMOS
image (epoch 2013) in Fig. 1 of \citet{Luhman13}. The fact that the
binary appears to have similar PA in 1984 and 2013 may be hinting that
the original estimate for the orbital period by \citet{Luhman13} of
$\sim$30 yr may not be far from the actual period. \citet{Hambly01}
estimate the 1$\sigma$ uncertainties in positions of stars on the
ESO-R plates to be 0''.133 in $\alpha$ and 0''.173 in $\delta$.\\

\begin{figure}[htb!]
\epsscale{1.0}
\plotone{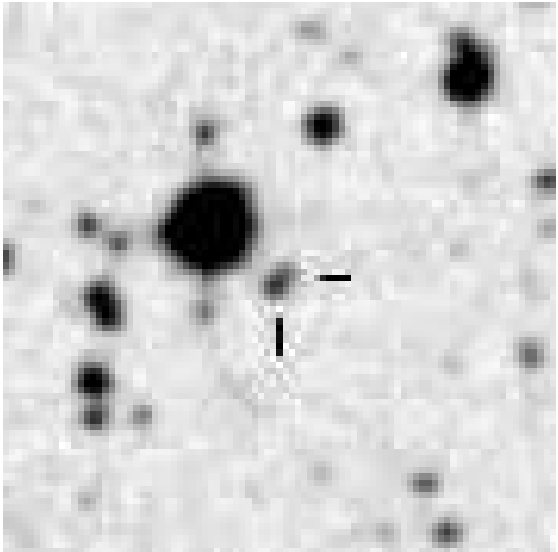}
\caption{Image of WISE J104915.57-531906.1 in SuperCOSMOS scan of red
  (IIIaF emulsion, RG630 filter) image taken with ESO Schmidt
  Telescope (epoch 1984.169). Field of view is 1 arcminute. Nothing is
  visible at the position of the elongated source on images taken at
  other epochs (compare with Fig. 1 of Luhman 2013).
\label{fig:ESO}}
\end{figure}

{\it AAO/UKST H$\alpha$ Survey (1999):}

\subsection{Summary on New Astrometry}

The new pre-WISE astrometry for WISE J104915.57-531906.1 is compiled in
Table \ref{tab:astrom}. Unfortunately the positional errors are
sizeable compared to the WISE, 2MASS, and DENIS astrometry already
presented in \citet{Luhman13}, and as astrometric solutions are
usually weighted by the inverse variance of the positional errors, I
have not bothered to recalculate an astrometric solution at this time.\\

\begin{deluxetable*}{llllll}[htb!]
\tabletypesize{\scriptsize}
\setlength{\tabcolsep}{0.03in}
\tablewidth{0pt}
\tablecaption{Additional Astrometry for WISE
J104915.57-531906.1\label{tab:astrom}}
\tablehead{
{(1)}   &{(2)}            & {(3)}            &{(4)}             &{(5)}           &{(6)}\\
{Alias or Survey} &{$\alpha_{ICRS}$} & {$\delta_{ICRS}$} &{$\sigma_{\alpha}$}&{$\sigma_{\delta}$} &Epoch        
}
\startdata
AKARI/IRC J1049166-531907 & 10:49:16.61 & -53:19:07.1 & 0''.21 & 0''.45  & 2007.0\\
...                       & 162.31920   & -53.31864   & ...    & ...     & ...\\
GSC2.3 S4BM006703 & 10:49:20.156 & -53:19:11.02 & 0''.419" & 0''.399" & 1995.304\\
...                       & 162.333984  & -53.319728  & ...    & ...  & ...\\
IRAS Z10473-5303  & 10:49:24.6   & -53:19:16    & 16''.5   & 5''.6    & 1983.5\\
ESO Schmidt (Red) & 10:49:23.7035 & -53:19:16.227 & 0''.2  & 0''.2 & 1984.169
\enddata
\tablecomments{The original AKARI/IRC error ellipse is 0''.45 $\times$
  0''.21 at PA = 89$^{\circ}$.66, and the original IRAS error ellipse
  is 16''.6 $\times$ 5''.2 at PA = 144$^{\circ}$. The original AKARI
  and GSC2.3 positions were quoted in decimal degrees, and the
  original IRAS and ESO Schmidt positions are provided in sexagesimal
  format.}
\end{deluxetable*}

\section{Kinematics}

\subsection{Thin Disk? Thick Disk? Halo?}

Even with just a proper motion and parallax, one can get a handle on
which Galactic kinematic population the brown dwarf likely belongs to.
The star's proper motion and parallax from \citet{Luhman13} are
consistent with a tangential velocity \vtan\, = 26.6\,$\pm$\,2.0 \kms,
which is typical for L dwarfs. \citet{Schmidt10} showed that the
distribution of tangential velocities for 748 spectroscopically
identified L dwarfs had median \vtan\, = 28 \kms\, with dispersion
$\sigma_{V_{tan}}$ = 25 \kms. The kinematic age of nearby L dwarfs in
the solar neighborhood is $\sim$3-5 Gyr \citep{Seifahrt10}, depending
on how the age is calculated based on the velocity distribution.  The
metallicities and ages of field L dwarfs are largely unknown, so
measurements of the velocity ellipsoid moments and their age evolution
are not well constrained. {\it If} we assume that the solar
neighborhood L dwarfs have velocity velocity distributions and
observed fractions of thin disk/thick disk/halo stars similar to that
of local G dwarfs \citep[adopted from Table 1 of ][]{Bensby03}, then
we can generate synthetic UVW velocities for thin disk/thick disk/halo
stars (and sampling within the parallax errors), and project the
synthetic velocities at the synethetic distance to produce synthetic
proper motions. Obviously this assumption can be criticized as one
expects differences between the kinematics and age distribution of L
dwarfs vs. G dwarfs, as the L dwarfs are expected to cool
significantly as they age, however the differences between the
kinematics of L and G dwarfs appear to be rather subtle (compare
e.g. \citet{Bensby03} with \citet{Seifahrt10} and
\citet{Schmidt10}).\\

I generated 10$^7$ synthetic thin disk/thick disk/halo stars
using the normalizations and velocity ellipsoids from
\citet{Bensby03}. Among the 10$^7$ synthetic stars with proper motion
vectors within 100 \masyr\, of the \pmra, \pmdec\, pair estimated by
\citet{Luhman13}, the simulation produced 377 thin disk stars (96\%),
16 thick disk stars (4\%), and zero halo stars. For the synthetic thin
disk stars, the predicted radial velocities had mean \vrad, = 0 \kms,
with rms = 16 \kms. For the thick disk stars, the predicted radial
velocities had mean \vrad\, = -18 \kms, with rms = 31 \kms. Hence, the
simulations predict that radial velocity of the system will likely not
be far from zero.\\

The high likelihood of the membership of this star to the Galactic
thin disk population (96\%) is not unexpected. \citet{Luhman13} has
already shown that the 2MASS and WISE infrared colors, and the
0.67-1.0 $\mu$m spectrum, appear to rather typical for a late L dwarf
in the solar neighborhood.\\

\subsection{Could This Binary Belong to any Nearby Moving Groups?}

At the time of writing, the proper motion and parallax for \name\,
have been measured \citep{Luhman13}, but a radial velocity has not yet
been measured (and given that the system is a binary with period
$>$decade, it is unlikely that a precise systemic radial velocity will
be known for some time, however rough estimates should be
forthcoming). With an accurate position and proper motion, one can
test whether the binary is moving towards the convergent point of
nearby moving groups.  Such calculations have been useful in testing
membership of young stars to nearby young moving groups
\citep[e.g.][]{Mamajek02, Mamajek05}.  Proper motions alone, however,
are insufficient for assigning group memberships. Typically when
radial velocities and trigonometric parallaxes are unavailable,
color-magnitude criterion can be used \citep[e.g.][]{Hoogerwerf00}.
Since a trigonometric parallax has been measured \citep{Luhman13}, one
can test whether the predicted cluster parallax (derived via the
proper motion and predicted tangential velocity) and measured parallax
agree. More sophisticated techniques have been developed within a
Bayesian framework \citep{Malo13}. However, I stick to the classic
convergent point methodology here for simplicity, as one can grasp the
degree of membership likelihood (or disagreement) by visually scanning
the quantities \pmtau\, (the proper motion component perpendicular to
the great circle between the star and the group convergent point;
which should be statistically consistent with zero for an ideal group
member), \vpec\, (the ``peculiar'' velocity; the tangential velocity
of the \pmtau\, motion, which similarly should be statistically
consistent with zero for an ideal member), and the predicted cluster
parallax (which in this case can be compared to the measured
trigonometric parallax). For a given group, if \pmtau\, and \vpec\,
are significantly non-zero, and/or the predicted cluster parallax does
not agree with the trigonometric parallax, then kinematic membership
to a stellar group can be ruled out.\\

The author maintains an as-yet unpublished compendium of kinematic,
position, age data for nearby young moving groups (clusters and
associations)\footnote{Available upon request.}. Some new kinematic
values in the database were presented in a poster on the kinematics of
young groups within 100 pc of the Sun using revised Hipparcos
astrometry\footnote{From \citet{vanLeeuwen07}.}  \citep{Mamajek10},
and in \citet{Mamajek08}, \citet{Luhman09}, \citet{Chen11},
\citet{Faherty13}, \citet{Barenfeld13}. Published kinematic data on
nearby groups can also be found in papers by \citet{deBruijne99,
  deBruijne01, Madsen02, Zuckerman04, Torres08, Malo13}. The adopted
convergent points, velocities, and 1D velocity dispersions for the
known groups within 100 pc are compiled in Table \ref{tab:groups}.\\

Using the convergent point method \citep[following][and references
  therein]{Mamajek02,Mamajek05}, and the convergent point solutions
for the nearby young stellar groups, for each group I rotated the
proper motion components \pmra\, and \pmdec\, for \name\, from
\citet{Luhman13} into the proper motion component towards the
convergent point (\pmups) and perpendicular to that component (\pmtau)
(see Table \ref{tab:cvp}). Using the \pmtau\, component and predicted
cluster parallax, one can also calculate a ``peculiar velocity'',
which puts an estimate on the minimum velocity difference between the
star's velocity and that of the group. As seen in Table \ref{tab:cvp},
none of the stellar groups provides a great combination of low
\pmtau\, and \vpec\, along with cluster parallax that agrees with the
observed trigonometric parallax. The most intriguing match is the
$\sim$40 Myr-old Argus group, which sports a somewhat low peculiar
velocity ($\sim$3 \kms), and a predicted parallax ($\varpi$\, =
510\,$\pm$\,25 mas) that isn't that far off (0.3$\sigma$) from the
trigonometric parallax ($\varpi$\, = 496\,$\pm$\,37 mas). If the object
belongs to the Argus group, it should have a radial velocity near
\vrad\, $\simeq$ 7.6 \kms. However, such a young age ($\sim$40 Myr)
seems unlikely given the near-IR colors of the binary.  The observed
$J-K_s$ color (1.89) is almost identical to the average color for L8/9
dwarfs \citep[1.85, rms = 0.17 mag; ][]{Faherty13}, and the observed
$W1 - W2$ color (0.56) is also almost identical to the average color
for L8/9 dwarfs \citep[0.54, rms = 0.08 mag;][]{Faherty13}.  The
observed $J-K_s$ and $W1-W2$ colors for $<$10$^8$ yr-old, low surface
gravity L dwarfs are systematically redder than those of older field L
dwarfs \citep[e.g.][]{Cruz09, Gizis12, Faherty13}, hence there is no
supporting evidence from the photometry to suggest that the binary
could be as young as the Argus group.\\

\begin{deluxetable*}{lrrcrc}[htb!]
\tabletypesize{\scriptsize}
\setlength{\tabcolsep}{0.03in}
\tablewidth{0pt}
\tablecaption{Adopted Convergent Point Solutions\label{tab:groups}}
\tablehead{
{(1)}   &{(2)}           & {(3)}           &{(4)}  & {(5)}  & {(6)}\\
{Group} & {$\alpha_{cvp}$} & {$\delta_{cvp}$} & {$S$} & {Ref.} & {\vdisp}\\
{}      & {deg}           & {deg}           & {\kms} & {} & {\kms}
}
\startdata
$\beta$ Pic group  &  87.9\,$\pm$\,0.9 & -30.5\,$\pm$\,0.8 & 21.4\,$\pm$\,0.2 & EEM                 & 1.5\\
AB Dor group       &  90.4\,$\pm$\,1.6 & -47.3\,$\pm$\,1.3 & 32.0\,$\pm$\,1.0 & \citet{Barenfeld13} & 1.0\\
UMa cluster        & 300.9\,$\pm$\,3.2 & -31.0\,$\pm$\,2.3 & 17.3\,$\pm$\,0.6 & \citet{Mamajek10}   & 1.3\\
Car-Near group     & 100.6\,$\pm$\,0.9 & -4.9\,$\pm$\,1.2  & 31.2\,$\pm$\,1.2 & EEM                 & 1.3\\
Hyades cluster     &  97.3\,$\pm$\,0.2 &  6.9\,$\pm$\,0.2  & 46.4\,$\pm$\,0.1 & \citet{deBruijne01} & 0.3\\
Tucana group       & 117.7\,$\pm$\,2.5 & -29.9\,$\pm$\,2.7 & 23.7\,$\pm$\,1.5 & EEM                 & 1.1\\
Argus group        &  91.9             & -1.2              & 25.3             & \citet{Malo13}      & 1.3\\
Columba group      & 103.3             & -29.9             & 25.2             & \citet{Malo13}      & 1.0\\
TW Hya group       & 101.5\,$\pm$\,0.6 & -29.4\,$\pm$\,0.5 & 21.9\,$\pm$\,0.2 & \citet{Weinberger13} & 0.8\\
Coma Ber cluster   & 114.5\,$\pm$\,1.8 & -34.5\,$\pm$\,1.4 & 6.0\,$\pm$\,0.1  & EEM                  & 1\\
32 Ori group       & 92.2\,$\pm$\,0.9  & -30.9\,$\pm$\,0.9 & 23.7\,$\pm$\,0.4 & EEM                  & 1\\
$\eta$ Cha cluster & 89.9\,$\pm$\,0.3  & -37.6\,$\pm$\,0.3 & 25.7\,$\pm$\,0.2 & \citet{Murphy10}     & 1\\
Alessi 13 cluster  & 104.5\,$\pm$\,2.7 & -28.8\,$\pm$\,2.2 & 24.8\,$\pm$\,0.9 & EEM                  & 1
\enddata
\tablecomments{Convergent points $\alpha_{cvp}$, $\delta_{cvp}$ and
  space velocities $S$ were calculated using the $UVW$ velocities from
  the papers listed. 1D velocity dispersions \vdisp\, are from the
  cited references, except that I adopt fiducial values of \vdisp\, =
  1 \kms\, for the Coma Ber, $\eta$ Cha, and Alessi 13 clusters, and
  0.8 \kms\, for TW Hya group from \citet{Mamajek05}. The exact choice
  of adopted 1D velocity dispersion has negligible impact on the
  calculations, and no impact on the conclusions for the binary being studied.}
\end{deluxetable*}

\begin{deluxetable*}{lccccc}[htb!]
\tabletypesize{\scriptsize}
\setlength{\tabcolsep}{0.03in}
\tablewidth{0pt}
\tablecaption{Convergent Point Calculations for \name \label{tab:cvp}}
\tablehead{
{(1)}   &{(2)}    & {(3)}    &{(4)}          & {(5)} & {(6)}\\
{Group} &{\pmups} & {\pmtau} &{$\varpi_{pred}$}& {\vpec} & {Member?}\\
{}      &{\masyr} & {\masyr} &{mas}          & {\kms} & {}        
}
\startdata
$\beta$ Pic group &  2683 &  -736 & 708\,$\pm$\,52 &  -4.9 & No\\
AB Dor group      &  2437 & -1342 & 497\,$\pm$\,23 & -12.8 & No\\
UMa cluster       & -1858 & -2071 & -510\,$\pm$\,42 & 19.2 & No\\
Car-Near group    &  2709 &   630 & 440\,$\pm$\,25  & 6.8  & No\\
Hyades cluster    &  2660 &   815 & 278\,$\pm$\,2   & 14.0 & No\\
Tucana group      &  2734 &   513 & 856\,$\pm$\,80  & 2.8  & No\\ 
Argus group       &  2762 &   329 & 510\,$\pm$\,25  & 3.1  & No?\\
Columba group     &  2779 &  -130 & 700\,$\pm$\,49  & -0.9 & No\\
TW Hya group      &  2776 &  -180 & 758\,$\pm$\,31  & -1.1 & No\\
Coma Ber cluster  &  2657 &  -823 & 3022\,$\pm$\,230 & -1.3 & No\\
32 Ori group      &  2716 &  -603 & 671\,$\pm$\,28  & -4.3  & No\\
$\eta$ Cha cluster & 2611 &  -959 & 622\,$\pm$\,25  & -7.3 & No\\
Alessi 13 cluster  & 2781 &  -34  & 713\,$\pm$\,47  & -0.2 & No
\enddata
\tablecomments{\pmups\, is the proper motion projected towards the
  group's convergent point, and \pmups\, is the perpendicular
  component. Uncertainties in both \pmups\, and \pmups\, are 6 \masyr,
  following \citet{Luhman13}.  $\varpi_{pred}$ is the predicted
  cluster parallax in mas, and \vpec\, is the predicted peculiar
  velocity. Negative \pmups\, and $\varpi_{pred}$ means that the star is
  moving in the opposite direction predicted for group members (hence,
  unphysically negative parallaxes!). \citet{Luhman13} calculated a
  trigonometric parallax of 496\,$\pm$\,37 mas.}
\end{deluxetable*}

\section{Comments on Nomenclature}

The name WISE J104915.57-531906.1 obviously follows IAU convention
(i.e. it is useful, conveys the position of the object in WISE images
at epoch ~2010, and "J" designates the equinox J2000 for the
equatorial coordinate system), but it is rather unglamorous and
lengthy for what is clearly a "special" pair of celestial objects that
will be heavily studied in the future. Given how important this system
will become given its provisional status as the nearest known system
of substellar objects and the third nearest "star system" (and I use
the term "star" in the general sense in that the pair is
self-luminous, but the objects are not "stars" in the sense that they
are hydrogen burning), a shorter name is warranted. ``Phone number''
names (e.g. WISE J104915.57-531906.1 or the shortened WISE J1049-5319
or even WISE 1049-5319) are fine for otherwise anonymous stars and
galaxies, but this pair of objects is special, and it seems silly to
call this object by a 24-character name (space included). Shortening
the WISE name to shorter versions (e.g. "WISE J1049", "WISE 1049",
``WISE 1049-5319'', ``WISE 1049-53'') may lead to confusion if other
interesting WISE-discovered objects are discovered in its vicinity
(something that occurred with shorter names used for 2MASS and SDSS
objects).\\

Luhman has already published several new binary star discoveries which
are compiled in the Washington Double Star
(WDS)\footnote{http://ad.usno.navy.mil/wds/} catalog with discovery
identifier "LUH". The WDS was originally published by \citet{Mason01},
and is updated frequently\footnote{The latest version of the WDS at
  the time of writing was dated 10 Mar 2013 and contained 125,271
  binaries.}. 15 "LUH" binaries are listed in the latest version of
WDS. B. Mason (private communication, 17 Mar 2013) has confirmed that
WISE J104915.57-531906.1 will be added to the WDS with discoverer
identification "LUH 16", i.e.  "Luhman 16".  Hence the components
could be called "Luhman 16A" and "Luhman 16B". As there will likely be
additional interesting nearby WISE objects found, an alternative idea
is to conjoin the names of the discoverer with ``WISE'':
i.e. ``Luhman-WISE 1''.  Either would be much more memorable than WISE
J104915.57-531906.1 and any shorthand variant (and confusion will be
avoided should another interesting WISE object be found with similar
RA).\\

\acknowledgements

I thank Brian Mason for comments on the WDS, and Valentin Ivanov,
Julian Girard, David Frew, and Kevin Luhman for discussions.

\vspace{1cm}

\noindent History:\\

\noindent 23 March 2013: minor edits to abstract. Replaced all
instances of the shortened name ``WISE J104915.57-531906'' with the
correct full name ``WISE J104915.57-531906.1''.\\

\end{document}